\documentclass{andromedaone}

\usepackage{amsmath,amssymb}

\journal{BSM}
\vol{2021}
\jyear{Egypt}
\pages{Hurgada} 

\def\be{\begin{equation}}
\def\ee{\end{equation}}
\def\bea{\begin{eqnarray}}
\def\eea{\end{eqnarray}}

\newcommand{\doublet}[2]{ \left( \begin{array}{c}#1 \\ #2 \end{array}\right) }

\begin{document}

\title{Baryogenesis from a CP-violating inflation}

\author{Venus Keus}
\address{\emph{\small  Department of Physics, University of Helsinki,}\\
\emph{\small P.O.Box 64, FI-00014 Helsinki, Finland Helsinki, Finland}}

\begin{abstract}
We introduce the novel phenomena of CP-violating inflation in the frameworks of a 3-Higgs doublet model where the inflaton doublets have a non-minimal coupling to gravity.
We allow for this coupling to be complex, thereby introducing CP-violation
- a necessary source of the baryon asymmetry - in the inflaton couplings.
We investigate the inflationary dynamics of such a framework and
the inflaton decay in the reheating phase.
We discuss how the CP-violation of the model is imprinted on the particle asymmetries.
\end{abstract}

\maketitle


\section{Introduction}

The Standard Model (SM) of particle physics has been extensively tested and is in great agreement with experimental data, with
its last missing particle – the Higgs boson – discovered by ATLAS and CMS experiments at the CERN Large Hadron Collider (LHC) \cite{Aad:2012tfa,Chatrchyan:2012ufa}.
Although the properties of the observed scalar are in agreement with those of the SM-Higgs boson, it may just be one member of an extended scalar sector.
Even though so far no signs of
new physics have been detected, it is well understood that the SM of particle physics is incomplete.

Cosmological and astrophysical observations imply a large dark matter (DM) component
in the energy budget of the universe.
Within the particle physics setting, this would be a particle which is stable on cosmological time scales, cold, non-baryonic, neutral and weakly interacting \cite{Ade:2015xua}.
A particle with such characteristics does not exist in the SM.
Another shortcoming of the SM is the lack of an explanation for the origin of the observed matter-antimatter asymmetry in the universe.
One of the most promising baryogenesis scenarios is electroweak baryogenesis (EWBG) \cite{Morrissey:2012db}, which produces the baryon excess during the electroweak phase transition (EWPT). Although the SM in principle contains all required ingredients for EWBG, it is unable to explain the observed baryon excess due to its insufficient amount of CP-violation \cite{Gavela:1993ts,Huet:1994jb,Gavela:1994dt} and the lack of a first-order phase transition \cite{Kajantie:1995kf}.

Furthermore, in its current form, the SM fails to incorporate inflation
in a satisfactory manner.
Inflation is a well-motivated theory predicting a period of exponential expansion
in the early universe
which explains the generation of primordial density fluctuations seeding structure formation,
flatness, homogeneity and isotropy of the universe
\cite{Hawking:1982cz,Starobinsky:1982ee,Sasaki:1986hm,Mukhanov:1988jd}.
The simplest models of inflation in best agreement with observations are those driven by a
scalar field, the \textit{inflaton}, with a standard kinetic term, slowly rolling down its
smooth potential.
At the end of inflation, the inflaton which naturally is assumed to have couplings with the SM-
Higgs, dumps its energy into the SM bath during the \textit{reheating} process which populates
the universe with SM particles.

Scalars with non-minimal couplings to gravity
are well-motivated inflaton candidates since they acquire fluctuations proportional to the inflationary scale and can drive the inflation process in the early universe,
as in the Higgs-inflation model \cite{Bezrukov:2007ep} where
the SM-Higgs plays the role of the inflaton, and $s$-inflation models \cite{Lerner:2009xg,Enqvist:2014zqa} where the SM is extended by a singlet scalar.
Extensive studies have been carried out in simple one singlet or one doublet scalar extensions of the SM (see e.g. \cite{Gong:2012ri,Choubey:2017hsq,Englert:2011yb,Branco:2011iw} and references therein).
These models, however, by construction can only partly provide a solution to the
main drawbacks of the SM.
For example, to incorporate both CP-violation and DM into the model
one has to go beyond simple scalar extensions of the SM \cite{Keus:2016orl};
see also e.g. \cite{Cordero-Cid:2016krd, Cordero:2017owj,Cordero-Cid:2018man,Cordero-Cid:2020yba,Keus:2020ooy,Keus:2019szx,Hernandez-Sanchez:2020aop}.

It is therefore theoretically appealing to have a more coherent setting where
different motivations of beyond SM (BSM) frameworks could be simultaneously investigated.
For example, in non-minimal Higgs frameworks with conserved discrete symmetries one
can accommodate stabilised DM candidates. Moreover, the extended scalar potential
could provide new sources of CP-violation and accommodate a strong first order
phase transition \cite{Ahriche:2015mea}. Collider searches can constrain these model frameworks by
excluding or discovering the existence of the spectrum of new states.

Here we introduce a model where a source of CP-violation
originates from the couplings of the inflation.
Through the process of reheating this is transmitted to an asymmetry within the
SM and can furthermore seed
the generation of an excess of matter over antimatter during the
evolution of the early universe.
We describe these dynamics in the context of a $Z_2$ symmetric
3-Higgs Doublet Model (3HDM) with a CP-violating extended dark sector,
which also provides a viable DM candidate, new sources of CP-violation and
a strong first-order EWPT \cite{Cordero-Cid:2016krd,Keus:2016orl,Cordero:2017owj,Cordero-Cid:2018man,Cordero-Cid:2020yba,Keus:2020ooy,Keus:2019szx,Hernandez-Sanchez:2020aop}.
We study the inflationary dynamics of this set-up and outline its main
consequences. 
We point out that the inflationary potential allows for very small scalar couplings of $\mathcal{O}(10^{-10})$ in agreement with all theoretical and experimental bounds, which consequently lead to a non-minimal coupling of $\xi \simeq 0.1$. Different values of the CP-violating angles then comfortably yield the conformal value of $|\xi|=1/6$.  
The thorough analysis of EWBG and DM observables as well as a phenomenological analysis
towards LHC searches of the model are covered in our upcoming publication.

\section{The scalar potential}
\label{sec:potential}
\subsection{General definitions}
A 3HDM scalar potential which is symmetric under a group $G$ of phase rotations, can be written as the sum of two parts: $V_0$ with terms symmetric under any phase rotation, and $V_G$ with terms symmetric under $G$ \cite{Ivanov:2011ae,Keus:2013hya}. As a result, a $Z_2$-symmetric 3HDM can be written as\footnote{We ignore additional $Z_2$-symmetric terms that can be added to the potential, e.g.,
$
(\phi_3^\dagger\phi_1)(\phi_2^\dagger\phi_3),
$
$
(\phi_1^\dagger\phi_2)(\phi_3^\dagger\phi_3),
$
$
(\phi_1^\dagger\phi_2)(\phi_1^\dagger\phi_1)
$
and
$
(\phi_1^\dagger\phi_2)(\phi_2^\dagger\phi_2),
$
as they do not change the phenomenology of the model \cite{Cordero-Cid:2018man}.}:
\bea
\label{eq:V0-3HDM}
V&=&V_0+V_{Z_2}, \\
V_0 &=& - \mu^2_{1} (\phi_1^\dagger \phi_1) -\mu^2_2 (\phi_2^\dagger \phi_2) - \mu^2_3(\phi_3^\dagger \phi_3) 
+ \lambda_{11} (\phi_1^\dagger \phi_1)^2+ \lambda_{22} (\phi_2^\dagger \phi_2)^2  + \lambda_{33} (\phi_3^\dagger \phi_3)^2 \nonumber\\
&& + \lambda_{12}  (\phi_1^\dagger \phi_1)(\phi_2^\dagger \phi_2)
 + \lambda_{23}  (\phi_2^\dagger \phi_2)(\phi_3^\dagger \phi_3) + \lambda_{31} (\phi_3^\dagger \phi_3)(\phi_1^\dagger \phi_1) 
 + \lambda'_{12} (\phi_1^\dagger \phi_2)(\phi_2^\dagger \phi_1)
 + \lambda'_{23} (\phi_2^\dagger \phi_3)(\phi_3^\dagger \phi_2) + \lambda'_{31} (\phi_3^\dagger \phi_1)(\phi_1^\dagger \phi_3),  \nonumber\\
 V_{Z_2} &=& -\mu^2_{12}(\phi_1^\dagger\phi_2)+  \lambda_{1}(\phi_1^\dagger\phi_2)^2 + \lambda_2(\phi_2^\dagger\phi_3)^2 + \lambda_3(\phi_3^\dagger\phi_1)^2  + h.c. \nonumber
\eea
where the three Higgs doublets, $\phi_{1},\phi_2,\phi_3$, transform under the $Z_2$ group, respectively, as
\be
\label{eq:generator}
g_{Z_2}=  \mathrm{\rm diag}\left(-1, -1, +1 \right).
\ee
The parameters of the $V_0$ part of the potential are real by construction. We allow for the parameters of $V_{Z_2}$ to be complex, using the following notation throughout the paper
\be
\label{eq:complex-params}
\lambda_{j} = |\lambda_{j}| \, e^{i\, \theta_{j}} \quad (j = 1,2,3), \quad
\quad
\mbox{and}
\quad
\mu^2_{12} = |\mu^2_{12}| \, e^{i\, \theta_{12}}\,.
\ee
The composition of the doublets is as follows:
\be
\phi_1= \doublet{$\begin{scriptsize}$ H^+_1 $\end{scriptsize}$}{\frac{H_1+iA_1}{\sqrt{2}}},\quad
\phi_2= \doublet{$\begin{scriptsize}$ H^+_2 $\end{scriptsize}$}{\frac{H_2+iA_2}{\sqrt{2}}}, \quad
\phi_3= \doublet{$\begin{scriptsize}$ G^+ $\end{scriptsize}$}{\frac{v+h+iG^0}{\sqrt{2}}},
\label{explicit-fields}
\ee
where $\phi_1$ and $\phi_2$ are the $Z_2$-odd \textit{inert} doublets,
$\langle \phi_1 \rangle = \langle \phi_2 \rangle =0$, and $\phi_3$ is the one
$Z_2$-even \textit{active} doublet, which at low energy attains a vacuum expectation
value (VEV)
$\langle \phi_3 \rangle =v/$\begin{scriptsize}$ \sqrt{2} $\end{scriptsize} $ \neq 0$.
The doublet $\phi_3$ plays the role of the SM Higgs doublet, with $h$ being the
SM Higgs boson and $G^\pm,~ G^0$ the would-be Goldstone bosons.
Note that
according to the $Z_2$ generator in Eq.~(\ref{eq:generator})
the symmetry of the potential is respected by the vacuum
$(0,0,v/$\begin{scriptsize}$ \sqrt{2} $\end{scriptsize}$)$. In this paper we consider the scenario where
the components of the inert doublets act as inflation candidates and reheat the
universe at the end of inflation through their interactions with the SM-Higgs and
gauge bosons. Note that at the scales relevant for inflation we can take the VEV
of the active doublet to be zero, $\langle \phi_3 \rangle = 0$.

Furthermore, CP-violation is only introduced in the \textit{inert} sector which is forbidden
from mixing with the \textit{active} sector by the conservation of the $Z_2$ symmetry. As a
result, the amount of CP-violation is not limited by electric dipole moments \cite{Cordero-Cid:2016krd}.
The lightest particle amongst the CP-mixed neutral fields from the inert doublets is a
viable DM candidate and stable due to the unbroken $Z_2$ symmetry.
In this paper, we focus on the inflationary dynamics of the model and shall not discuss DM
implications of the model any further.

\subsection{Potential for the inflaton}

We start by rewriting the doublets in the unitary gauge and ignore the charged scalars
(since they do not affect the inflationary dynamics).
\be
\phi_1= \frac{1}{\sqrt{2}} \left(\begin{array}{c}
0 \\[2mm]
h_1 +i \eta_1
\end{array} \right)
,\quad
\phi_2= \frac{1}{\sqrt{2}} \left(\begin{array}{c}
0 \\[2mm]
h_2 + i \eta_2
\end{array} \right)
, \quad
\phi_3= \frac{1}{\sqrt{2}} \left(\begin{array}{c}
0 \\[2mm]
h_3
\end{array} \right).
\label{eq:explicit-fields}
\ee

The action of the model in the Jordan frame is
\bea
\label{Eq:action-Jordan}
S_J &=& \int d^4x \sqrt{-g}
\bigg[  -\frac{1}{2} M^2_{pl} R - D_\mu \phi_1^\dagger  D^\mu \phi_1
- D_\mu \phi_2^\dagger  D^\mu \phi_2
- D_\mu \phi_3^\dagger  D^\mu \phi_3
\\
&& \hspace{2.3cm}
- V(\phi_1, \phi_2,\phi_3)
- \biggl(\xi_1 |\phi_1|^2  +\xi_2 |\phi_2|^2 +\xi_3 |\phi_3|^2
+ \xi_4 (\phi^\dagger_1 \phi_2)  +\xi_4^* (\phi^\dagger_2 \phi_1 ) \biggr)R
\bigg],
\nonumber
\eea
where $R$ is the Ricci scalar, $M_{pl}$ is the reduced Planck mass and the parameters $\xi_i$ are dimensionless couplings of the scalar doublets to gravity.
Note that, in principle, $\xi_4$ could be a complex parameter for which we use the notation
$
\xi_4 = |\xi_4|\, e^{i\theta_4}$.

In Eq.~(\ref{Eq:action-Jordan}) the covariant derivative, $D_\mu$, contains couplings of the scalars with the gauge bosons. However, for the dynamics during the inflation,
the covariant derivative is reduced to the normal derivative
$D_\mu\rightarrow \partial_\mu$. The minus sign in the kinetic terms follows the metric convention of $(-,+,+,+)$.

Since we identify the two inert doublets with inflaton, we assume that the energy density of
$\phi_3$ is sub-dominant during inflation. Therefore, the
part of the potential relevant for inflation
is
\be 
\label{eq:approxscalarpot}
V  =  - \mu^2_{1} (\phi_1^\dagger \phi_1) -\mu^2_2 (\phi_2^\dagger \phi_2)
+ \lambda_{11} (\phi_1^\dagger \phi_1)^2+ \lambda_{22} (\phi_2^\dagger \phi_2)^2 
 + \lambda_{12}  (\phi_1^\dagger \phi_1)(\phi_2^\dagger \phi_2)
 + \lambda'_{12} (\phi_1^\dagger \phi_2)(\phi_2^\dagger \phi_1)
 -\mu^2_{12}(\phi_1^\dagger\phi_2)+  \lambda_{1}(\phi_1^\dagger\phi_2)^2 + h.c. 
\ee 
Due to local SU(2) invariance, we can rotate away one of the CP-odd fields,
say $\eta_2$.
Such a transformation is equivalent to taking the $\eta_2 \to 0$ limit, and
we assume this limit to be taken when writing the fields in terms
of components in Eq.~(\ref{eq:explicit-fields}).


To facilitate the analysis, we apply a conformal transformation from 
the Jordan frame, which contains terms with scalar-gravity quadratic couplings,
to the Einstein frame with no explicit couplings to gravity \cite{Kaiser:2010ps}.
Physical observables are invariant under this frame transformation. The two frames
are equivalent after the end of inflation when the transformation parameter equals unity.
The action in the Einstein frame can be written as
\be
S_E=\int d^4x \sqrt{-\tilde g}\left[-\frac{1}{2}M_{pl}^2\tilde R-\frac{1}{2}\tilde{g}^{\mu\nu}\, G_{ij} \,\partial_\mu \varphi_i \partial_\nu\varphi_j-\tilde{V}\right],
\label{eq:action-Einstein}
\ee
where $\tilde{V} = V/\Omega^4$ is the potential in the Einstein frame following
the conformal transformation
\be 
\tilde{g}_{\mu\nu}=
\Omega^2 g_{\mu\nu}\, ,
\qquad
G_{ij} =
\frac{1}{\Omega^2}\delta_{ij}+\frac{3}{2}\frac{M_{pl}^2}{\Omega^4}\frac{\partial\,\Omega^2}{\partial\,\varphi_i}\frac{\partial\,\Omega^2}{\partial\,\varphi_j},
\label{eq:conformal-transf}
\ee 
where $\varphi_k=h_{1},h_{2},\eta_{1}$, and the transformation parameter
\be
\label{eq:def-Omega}
\Omega^2 =
1 +\frac{\xi_1}{M_{pl}^2}(h_1^2+\eta_1^2)+\frac{\xi_2}{M_{pl}^2}h_2^2+\frac{2 |\xi_4|}{M_{pl}^2}
\biggl( h_1h_2 c_{\theta_4} +\eta_1 h_2 s_{\theta_4}\biggr)
\ee
using the shorthand notation $c_{\theta_k} = \cos\theta_k $ and $s_{\theta_k} = \sin\theta_k $ throughout the paper.

The prefactor $G_{ij}$ in Eq.~(\ref{eq:conformal-transf}) leads to mixed kinetic terms.
We introduce the reparametrisation
\be
\label{eq:def-A}
A =\sqrt{\frac{3}{2}} \, M_{pl} \, \log (\Omega^2)
\qquad \mbox{with} \qquad
\frac{\partial\,\Omega^2}{\partial\,\varphi_k} =
\sqrt{\frac{2}{3}}\, \frac{\Omega^2}{M_{pl}} \, \frac{dA}{d\varphi_k}
\ee
which reduces the kinetic terms to the diagonal form
\be
\label{eq:diag-kinetic}
\tilde{g}_{\mu\nu} \, G_{ij} \,  \partial_\mu \varphi_i \partial_\nu\varphi_j
=
\Omega^2 g_{\mu\nu} \biggl( \frac{\delta_{ij}}{\Omega^2}
+  \frac{\partial A}{\partial \varphi_i}  \, \frac{\partial A}{\partial \varphi_j} \biggr)
\partial_\mu \varphi_i \partial_\nu \varphi_j
\, = \,
\partial_\mu \varphi_i \partial_\mu \varphi_i
+ \Omega^2 \, \partial_\mu A \,  \partial_\mu A
\ee


To write the potential in the Einstein frame, 
we keep only terms in the potential in Eq.~(\ref{eq:approxscalarpot})
which are quartic in $h_{1,2}$ and $\eta_{1}$.
This reduces the potential to
\be 
\tilde{V}  \approx   \frac{1}{4 \, \Omega^4} \biggl[
\lambda_{11} (h_1^2 + \eta_1^2)^2
+\lambda_{22}  h_2^4
+(\lambda_{12}+\lambda'_{12})(h_1^2 + \eta_1^2)h_2^2
+ 2 |\lambda_1| \biggl( c_{\theta_1}
\left(h_2^2(h_1^2 - \eta_1^2) \right)
+ 2\, s_{\theta_1} h_2^2h_1\eta_1  \biggr)
 \biggr] 
\label{eq:tildeV-quartic}
\ee 
where $\theta_1$ is the CP-violating phase of the $\lambda_1$ parameter.


Further, we introduce the reparametrisation
$\eta_1 = \beta_1 \, h_1$ and $h_2 = \beta_2 \, h_1$
with $\beta_1,\beta_2$ as field dependent values, to rewrite the potential as
\be
\label{eq:pot-with-B1B2}
\tilde{V} \approx
\frac{h_1^4}{4 \, \Omega^4} \biggl[
\lambda_{11}  (1+ \beta_1^2)^2
+\lambda_{22} \,  \beta_2^4
+ \biggl(
(\lambda_{12}+\lambda'_{12})(1 + \beta_1^2)
+ 2 |\lambda_1| \left( c_{\theta_1}
(1 - \beta_1^2)
+ 2\, s_{\theta_1} \beta_1 \right)\biggr) \beta_2^2
 \biggr]
\ee
Using this reparametrisation, one can also simplify the $\Omega^2$ parameter in Eq.~(\ref{eq:def-Omega}) as
\be
\label{eq:def-OmegaSq}
\Omega^2 =
1+ \left(\frac{\xi_1}{M_{pl}^2}(1+\beta_1^2)+\frac{\xi_2}{M_{pl}^2}\beta_2^2+\frac{2 \,|\xi_4|}{M_{pl}^2}
\beta_2(c_{\theta_4} +\beta_1 s_{\theta_4})\right) \, h_1^2\equiv 1+\frac{B}{M_{pl}^2}h_1^2.
\ee
From Eq.~(\ref{eq:def-A}), recall that $\Omega^2= \exp(\tilde{A})$ using the shorthand notation
$\tilde{A}=\sqrt{\frac{2}{3}}\frac{A}{M_{pl}}$.
One can then write the field $h_1$ in terms of the reparametrised field $\tilde{A}$
\be
\label{eq:h1-A-relation}
h_1^2 =
\frac{M_{pl}^2}{B}
\left(e^{\tilde{A}} -1 \right)\,.
\ee
Therefore, expressing $h_1^2$ and $\Omega^2$ in terms of $\tilde{A}$ allows us to write the
potential in Eq.~(\ref{eq:pot-with-B1B2}) in the form
\be
\tilde{V}\sim (1-e^{-\tilde{A}})^2 X(\beta_1,\beta_2).
\ee

We will be interested in the effect of the non-minimal coupling $|\xi_4|$ and
the associated phase $\theta_4$. Therefore, we will set $\xi_1=\xi_2=0$
and assume that the initial field values are such that $\Omega^2>0$ is guaranteed.
Therefore, with these assumptions,
the potential in Eq.~(\ref{eq:pot-with-B1B2}) can be written as
\be
\label{eq:full-pot-B1B2}
\tilde{V}=
\biggl(\frac{M_{pl}^2}{2 \,|\xi_4|} \biggr)^2
\left(1-e^{-\tilde{A}}\right)^2 \,
X(\beta_1,\beta_2)\,
\ee
where
\begin{small}
\be
X(\beta_1,\beta_2) =
\frac{\lambda_{11}  (1+ \beta_1^2)^2
+\lambda_{22} \,  \beta_2^4
+ \left(
(\lambda_{12}+\lambda'_{12})(1 + \beta_1^2)
+ 2 |\lambda_1| \left( c_{\theta_1}
(1 - \beta_1^2)
+ 2\, s_{\theta_1} \beta_1 \right)\right) \beta_2^2 }
{4\beta_2^2(c_{\theta_4} + \beta_1\, s_{\theta_4})^2 } \,.
\ee
\end{small}
Following the procedure in \cite{Gong:2012ri,Keus:2021dti}, to find the direction of inflation, we minimise the $X(\beta_1,\beta_2)$ function with respect to $\beta_1$ and $\beta_2$ 
to find the form of the $X$ function independent of $\beta_1$ and $\beta_2$ with only $\theta_1$ and $\theta_4$ as variables (see \cite{Keus:2021dti} for detailed derivation):
\be
\label{eq:X-thetas}
X(\theta_1,\theta_4) = \frac{\frac{1}{4}\left(\lambda_{12}+\lambda'_{12}+2\sqrt{\lambda_{11}\lambda_{22}}\, \right)^2 - \lambda_1^2  }{\lambda_{12}+\lambda'_{12}+2\sqrt{\lambda_{11}\lambda_{22}} - 2\lambda_1 \cos(\theta_1-2\theta_4)}\,.
\ee

\section{Inflationary dynamics}
\label{sec:slow-roll}

\begin{figure}[t!]
\centering
\includegraphics[scale=0.8]{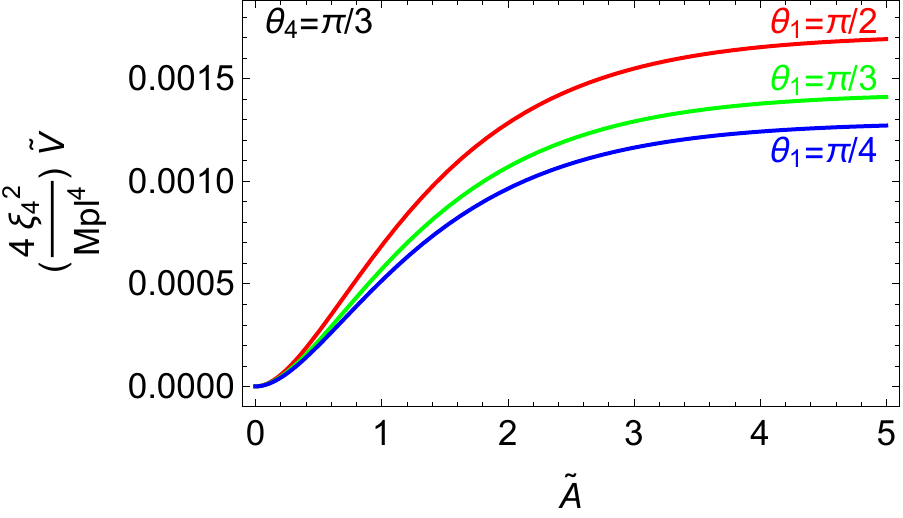}~~
\includegraphics[scale=0.8]{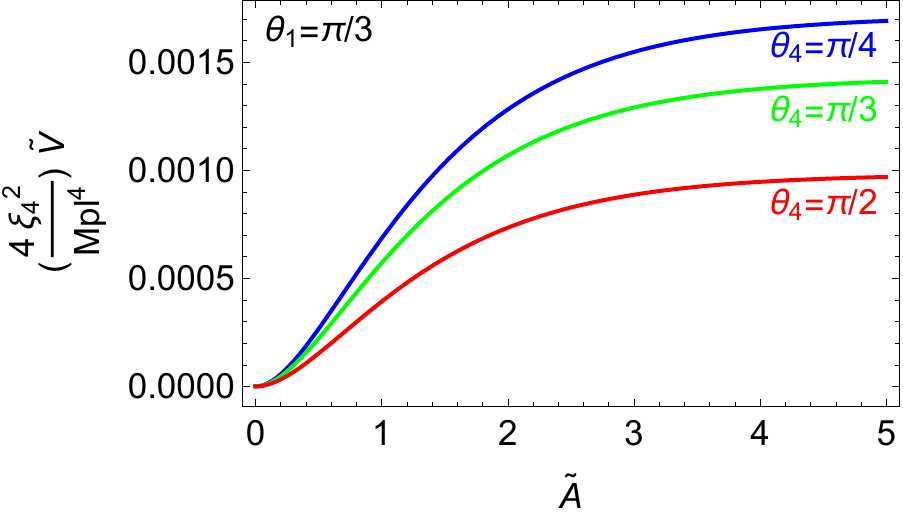}
\caption{The inflationary potential for different values of $\theta_1$ and $\theta_4$ (all $\lambda_i \sim 0.001$).}
\label{Fig:V-vs-ts}
\end{figure}

With the procedure used in the previous section,
the dynamics is essentially that of a single field inflation.
The full inflationary potential in Eq.~(\ref{eq:full-pot-B1B2}) can be written as
\be
\label{eq:full-pot-T1T4}
\tilde{V}= \biggl(\frac{M_{pl}^2}{2 \,|\xi_4|} \biggr)^2
\left(1-e^{-\tilde{A}}\right)^2 \,
X(\theta_1,\theta_4)\,
\ee
Figure \ref{Fig:V-vs-ts} shows the inflationary potential for different values of  $\theta_1$ and $\theta_4$ for a given value of $\lambda_i \sim 0.001$. Note that the potential is almost flat at high field values which ensures a slow roll inflation.

For the usual slow roll parameters in this case the function $X$ is irrelevant,
since it cancels in the expressions for $\epsilon$ and $\eta$, which are
\bea
\epsilon =
\frac{1}{2}M_{pl}^2 \,\left(\frac{1}{\tilde{V}}\frac{d\tilde{V}}{dA}\right)^2 =
\frac{4}{3\left( 1- e^{\tilde{A} }  \right)^2}\,,
\quad \qquad
\eta =
M_{pl}^2 \, \frac{1}{\tilde{V}}\frac{d^2\,\tilde{V}}{dA^2}
=
\frac{4(2-e^{\tilde{A}})  }{3\left( 1- e^{\tilde{A} }  \right)^2}\,.
\label{eta}
\eea
For field values $A\gg M_{pl}$ (or equivalently $\tilde{A}\gg 1 $), both parameters $\epsilon,\eta \ll 1$ which satisfies the slow roll condition. Inflation ends when $\epsilon \simeq 1$.
To calculate the values of $A$ at the beginning and end of inflation,  $A_i$ and $A_f$ respectively, one needs to calculate the number of e-folds $N_e$, i.e. the number of times the universe expanded by $e$ times its own size. $N_e$ is calculated to be
\bea
\label{eq:efolds}
N_e = \frac{1}{M_{pl}^2}\int_{A_{f}}^{A_{i}} \frac{\tilde{V}}{\tilde{V}'}\,dA
\, = \,
\frac{3}{4}
\left[ \tilde{A}_f - \tilde{A}_i
-e^{\tilde{A}_f} + e^{\tilde{A}_i}
\right] ,
\eea
where $\tilde{V}'=\frac{d\tilde{V}}{dA}$ and $A_{i}$ ($\tilde{A}_i$) is the value of $A$ ($\tilde{A}$) at the beginning of inflation and $A_{f}$  ($\tilde{A}_f$) is the value of $A$ ($\tilde{A}$) at the end of the inflation.
Since inflation ends when $\epsilon\simeq 1$, one can calculate $A_{f}$, which yields:
\be
e^{\tilde{A}_f} = \exp\left(\sqrt{\frac{2}{3}}\frac{A_{f}}{M_{pl}}\right)  \simeq  2.1547
\quad \Rightarrow \quad
\tilde{A}_f = \sqrt{\frac{2}{3}}\,\frac{A_{f}}{M_{pl}} \simeq
0.7676 \, .
\label{eq:Af}
\ee
To calculate $A_{i}$, one could plug in the $A_{f}$ value into Eq.~(\ref{eq:efolds}) assuming $N_e=60$, which results in
\be
\frac{3}{4}
\left[-\tilde{A}_i+  e^{\tilde{A}_i}
\right]- 1.0403
=60,
\quad
\Rightarrow  \quad
\tilde{A}_i = \sqrt{\frac{2}{3}}\frac{A_{i}}{M_{pl}}
\approx
4.4524
\ee


At this point we can also check
the field values in terms of the
original field $h_1$ using Eq.~(\ref{eq:h1-A-relation}).
This gives
\be
\label{eq:hi&hf}
 {h_1}_f  =
\frac{1.85 \times 10^{18}  }{\sqrt{|\xi_4|\,\beta_2(c_{\theta_4} + \beta_1\, s_{\theta_4}) }}\,,
\qquad
{h_1}_i  =
\frac{1.59 \times 10^{19} }{\sqrt{|\xi_4|\,\beta_2(c_{\theta_4} + \beta_1\, s_{\theta_4}) }} \,.
\ee

%

\begin{figure}[t!]
\centering
\includegraphics[scale=0.6]{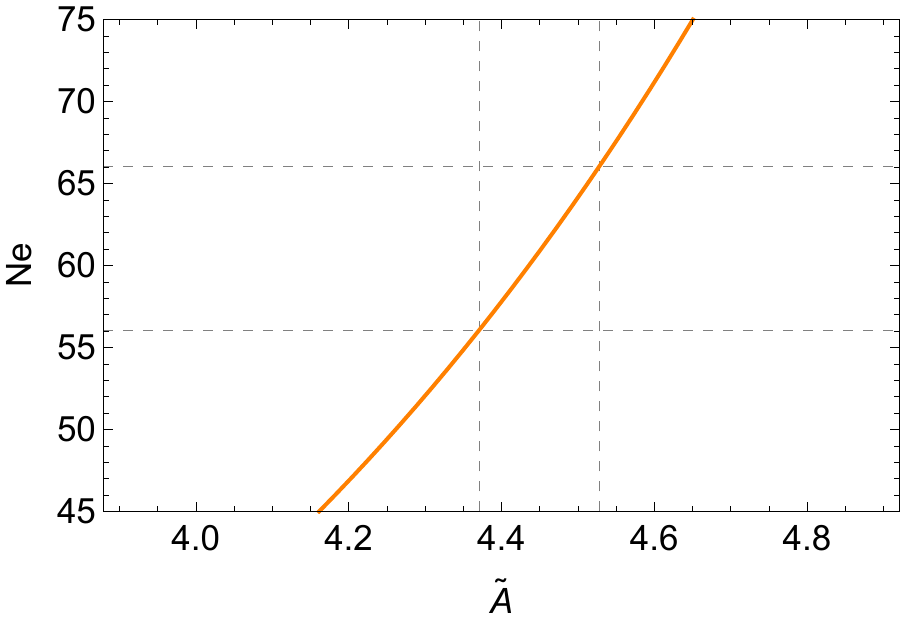}~
\includegraphics[scale=0.6]{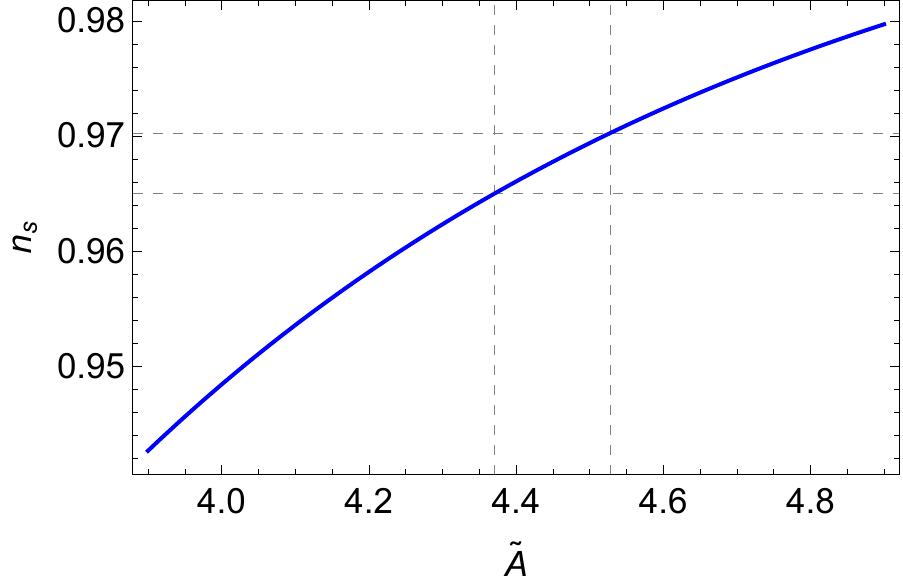}~
\includegraphics[scale=0.6]{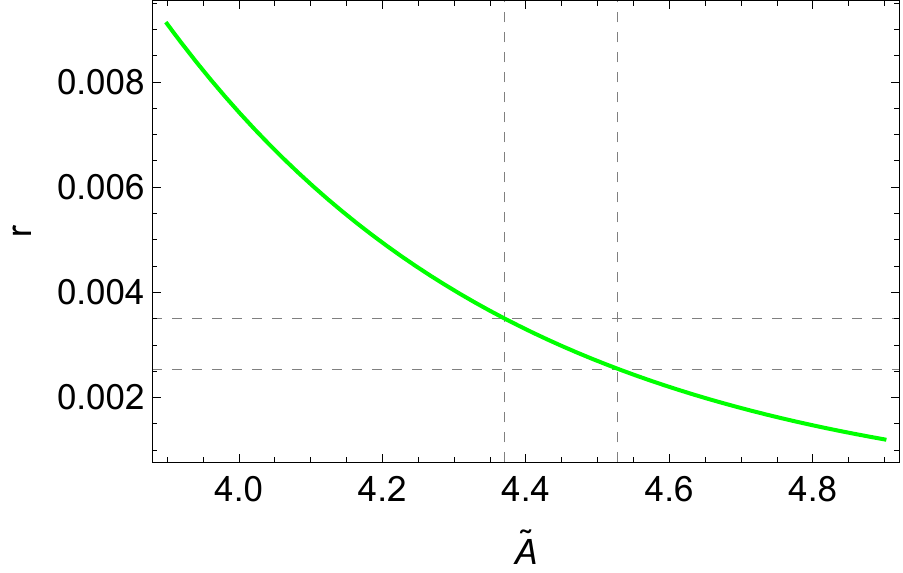}
\caption{The slow roll parameters: the number of $e$-folds $N_e$ (left), spectral index $n_s$ (center) and tensor to scalar ratio $r$ (right) as a function of $\tilde{A}$ with the grid-lines highlighting the $55 <N_e <65$ values.}
\label{Fig:Ps-r-ns}
\end{figure}
\begin{figure}[h!]
\centering
\includegraphics[scale=0.9]{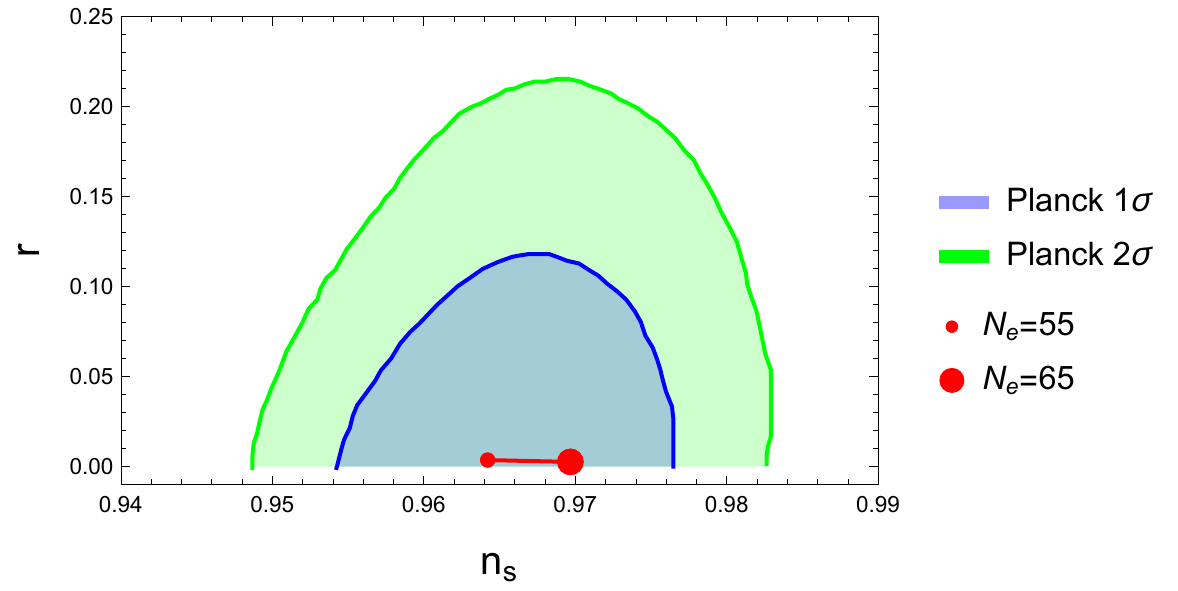}
\caption{The 1$\sigma$ and 2$\sigma$ regions for $n_s$ and $r$
from Planck observation compared to the theoretical
prediction of our framework.}
\label{Fig:Planck-nsVsr}
\end{figure}

Having fixed $N_e$ to 60, and calculated the $A$ field value at the start of inflation, we can derive the scalar power spectrum, $P_s$, the tensor to scalar ratio $r$ and the spectral index $n_s$ as follows:
\bea
P_s =\frac{1}{12\,\pi^2 \, M_{pl}^6} \frac{ \left({\tilde{V}}\right)^3}{\left({\tilde{V}'}\right)^2 }
&=&
\left(\frac{ (1-e^{\tilde{A}})^4}{128 \, \pi^2\, e^{2\tilde{A}}}\right)\frac{X(\theta_1,\theta_4)}{|\xi_4|^2}
=
5.565 \times \frac{X(\theta_1,\theta_4)}{|\xi_4|^2} ,
\\
\nonumber\\
r =16\,\epsilon &=& 0.00296,
 \\
n_s =1-6\epsilon+2\eta &=& 0.9678,
\eea
where $\tilde{V}'$ is the derivative of $\tilde{V}$ with respect to $A$ and both $\tilde{V}$ and $\tilde{V}'$ are calculated at the $A_{i}$.
Figure \ref{Fig:Ps-r-ns} shows the slow roll parameters $N_e$, $n_s$ and $r$ with respect to $\tilde{A}$ with the grid-lines highlighting the $55 <N_e <65$ values. We show the inflationary parameters over a range of $N_e$, since there is no reason for $N_e$ to be precisely $60$.
The values of $r$ and $n_s$ are well within the Plank bounds of $n_s=0.9677\pm 0.0060 $ at $1\sigma$ level and $r<0.11$ at $95\%$ confidence level \cite{Ade:2015lrj}. Note that the spectral index and the tensor to scalar ratio are in agreement with the Planck bounds over the full range of $N_e$.
Figure~\ref{Fig:Planck-nsVsr} shows the 1$\sigma$ and 2$\sigma$ regions allowed by Planck observations in the $r$-$n_s$ plane and   the theoretical predictions of our framework for $N_e$ values of 55 and 65.

\begin{figure}[t!]
\centering
\includegraphics[scale=0.7]{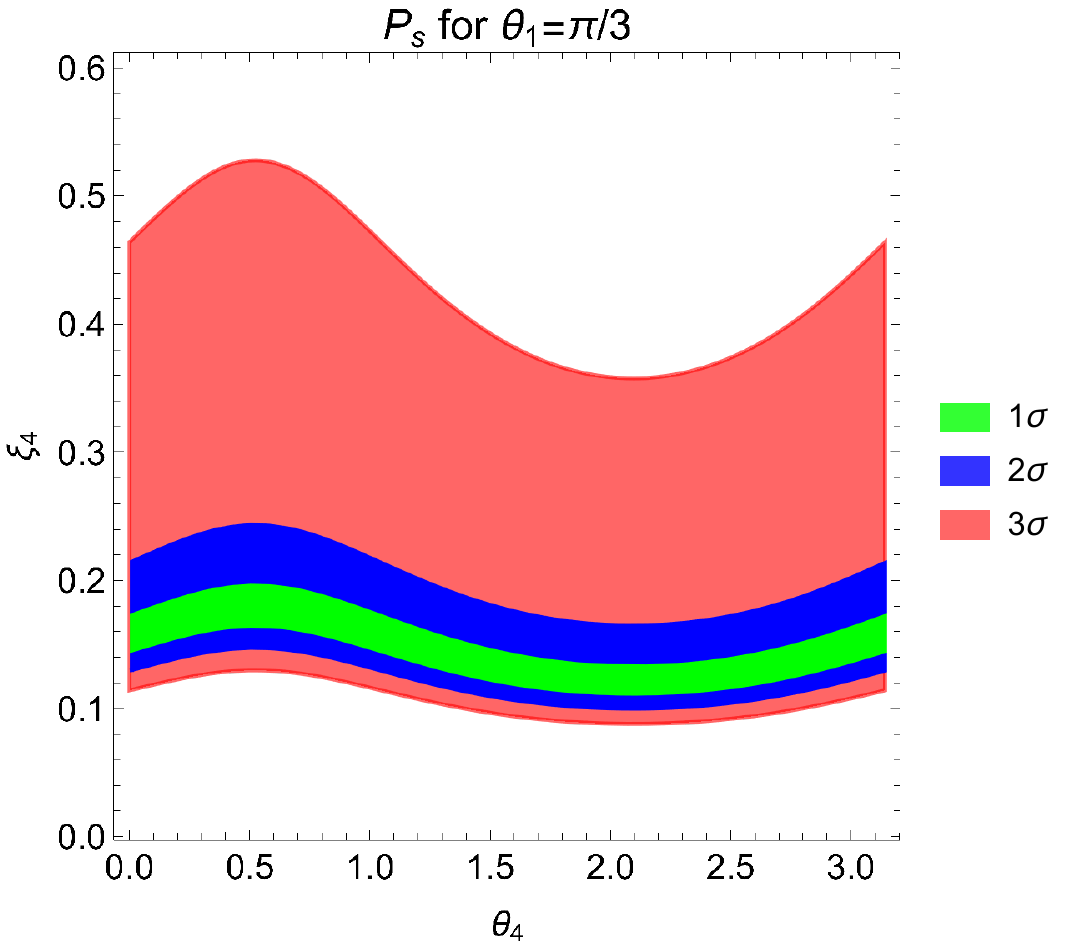}~
\includegraphics[scale=0.65]{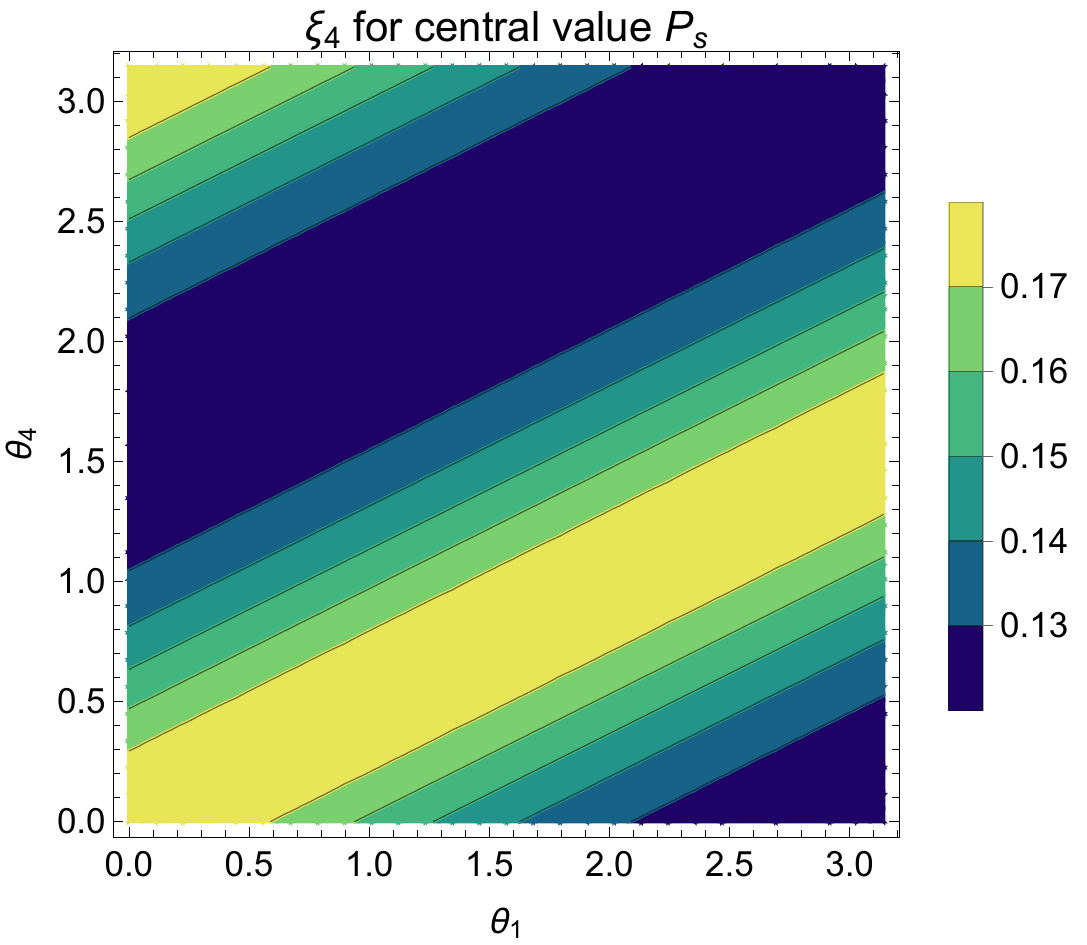}
\caption{Left panel: $P_s$ values for the fixed $\theta_1=\pi/3$ angle and varying values of $|\xi_4|$ and $\theta_4$ up to $3\sigma$ standard deviation from the observed central value. Right panel: Contours of $|\xi_4|$ in the $\theta_1$-$\theta_4$ plane which lead to $P_s$ central values (all $\lambda_i \sim \mathcal{O}(10^{-10})$).}
\label{Fig:Ps-Xi4}
\end{figure}

Observations from WMAP7 \cite{Komatsu:2010fb} constrain the scalar power spectrum which put a bound on the $|\xi_4|$  coupling and angles $\theta_1,\theta_4$,
\be
\label{eq:PS-bound}
P_s=(2.430\pm 0.091)\times 10^{-9}
\, = \,
5.565 \times \frac{X(\theta_1,\theta_4)}{|\xi_4|^2} \, .
\ee
In the left panel of Figure \ref{Fig:Ps-Xi4}, we show $P_s$ values for the fixed $\theta_1=\pi/3$ angle and varying values of $|\xi_4|$ and $\theta_4$ up to $3\sigma$ standard deviation from the central value in Eq.~(\ref{eq:PS-bound}). In the right panel, we fix $P_s$ to the WMAP7 central value for fixed values of $\lambda_i \sim 0.001$ to get
\be
|\xi_4|= 4.785\times 10^{4}\, \sqrt{X(\theta_1,\theta_4) }\,
\ee
and show contours of $|\xi_4|$ for varying values of $\theta_1$ and $\theta_4$.
Note that every point in the plot yields the exact $P_s$ central value.

This is a very important feature of our framework. To satisfy the bounds on the scalar power spectrum, the function $X(\theta_1,\theta_4)$ allows for a wide range of $\xi_4$ values as shown in Figure~\ref{Fig:Ps-Xi4}. This is in contrast to
the Higgs-inflation models where $P_s \propto {\lambda}/{\xi^2}$
with $\lambda$ the Higgs self-coupling which is fixed to be $\sim 0.12$ at the electroweak scale. Thus, for $P_s$ to agree with observations at the inflationary scale, $\xi$ will have to be very large $\mathcal{O}(10^4)$.
In our set-up, a combination of parameters $\lambda_{1}, \lambda_{11}, \lambda_{22}, \lambda_{12},\lambda'_{12}$ appears in the $X(\theta_1,\theta_4)$ function. The only constraint limiting these parameters is the stability of the potential requiring
\bea
\lambda_{ii} > 0, \qquad
\lambda_{ij} + \lambda'_{ij} > -2 \sqrt{\lambda_{ii}\lambda_{jj}}, \qquad
|\lambda_i| \leq |\lambda_{ii}|, |\lambda_{ij}|, |\lambda'_{ij}| , \quad i\neq j = 1,2,3\, ,
\eea
which allows for very small values of $\lambda_i \sim \mathcal{O}(10^{-10})$ which, in turn, allows for much smaller values of $|\xi_4|$ of order $0.1$. Different values of the CP-violating angles $\theta_1$ and $\theta_4$ could then comfortably yield the conformal value of $|\xi|=1/6$.  

\section{Reheating and inflaton decay}
\label{sec:reheating}

\begin{figure}[t!]
\begin{center}
\includegraphics[scale=0.9]{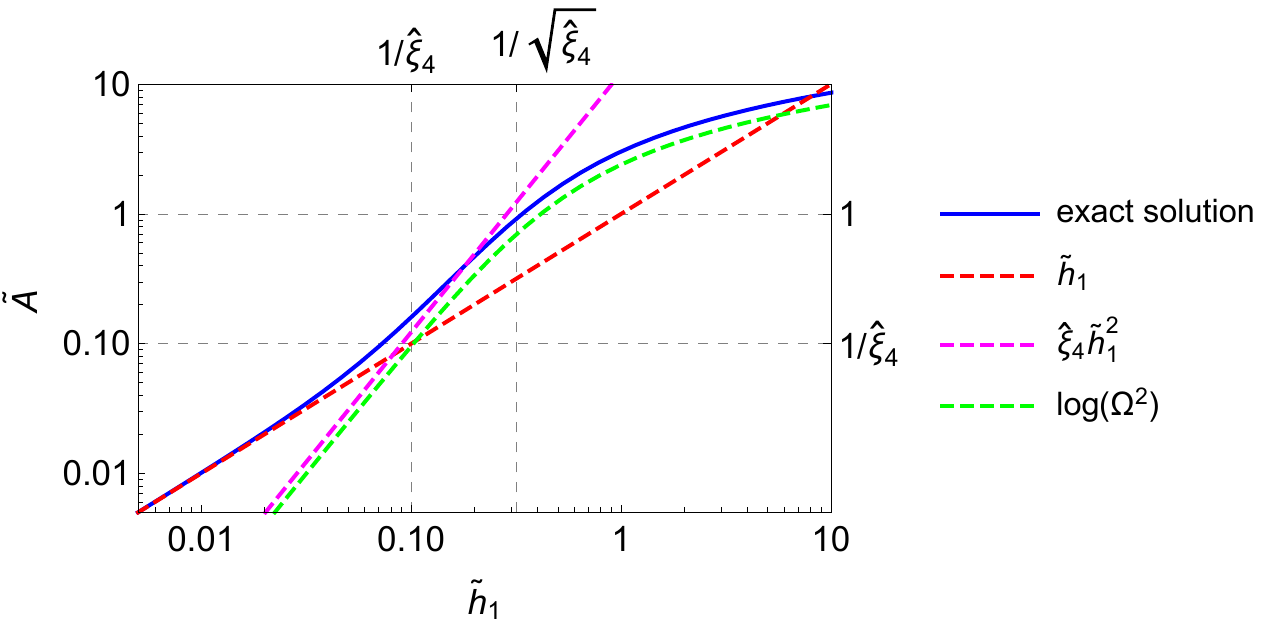}
\caption{The reparametrised field $\tilde{A}=\sqrt{\frac{2}{3}}\,A/M_{pl}$ with respect to the $\tilde{h}_1=h_1/M_{pl}$ field. Note that at low field values (end of inflation) $\tilde{A}$ and $\tilde{h}_1$ coincide as expected.}
\label{Fig:ALegendLogPlot}
\end{center}
\end{figure}

At the end of inflation, the energy stored in the inflaton disperses as the inflaton decays/annihilates into the SM particles through processes mediated by the SM-Higgs and gauge bosons in our case, during the so-called reheating phase \cite{Linde:1981mu}. 
To dissect the reheating phase of our CP-violating inflationary dynamics, we make use of the conformal transformation and field redefinitions in 
Eqs.~(\ref{eq:conformal-transf}-\ref{eq:def-OmegaSq}) to write 
\be 
\left(\frac{dA}{d h_1} \right)^2 = 
\frac{1}{\Omega^2} + \frac{1}{\Omega^4} \frac{24 }{M_{pl}^2}\,|\xi_4|^2 \beta_2^2(c_{\theta_4} +\beta_1 s_{\theta_4})^2 h_1^2 \,,
\ee
or simply
\be 
\label{eq:Atl-htl-integral}
\frac{d \tilde{A}}{d \tilde{h}_1} 
=
\frac{\sqrt{1+\hat{\xi}_4(1+6\hat{\xi}_4 )\, {\tilde{h}_1}^2}}{1+\hat{\xi}_4 {\tilde{h}_1}^2}\,,
\ee 
using the shorthand notations
\be 
\label{eq:shorthand}
\tilde{h}_1 = \frac{h_1}{M_{pl}}, \qquad
\hat{\xi}_4 =  2 \,|\xi_4|
\beta_2(c_{\theta_4} +\beta_1 s_{\theta_4})\, ,\qquad \mbox{and} \quad
\tilde{A}=\sqrt{\frac{2}{3}}\frac{A}{M_{pl}}\,,
\ee
as defined before. The exact solution to Eq.(\ref{eq:Atl-htl-integral}) after integration (whose analytical form is not particularly enlightening) is shown in Figure~\ref{Fig:ALegendLogPlot} represented by the solid blue curve. 
We find it instructive to identify two distinct regions as
\be 
\tilde{A} \approx 
\Bigg{\lbrace} 
\begin{array}{c} 
\tilde{h}_1  \hspace{19mm} \textrm{for }\tilde{A} < \tilde{A}_{cr}\, ,
\\[2mm] 
\log\left(\Omega^2\right) \qquad \textrm{for }\tilde{A}>\tilde{A}_{cr}\,,
\end{array}
\ee
where at the end of inflation (at low field values), the field $\tilde{h}_1$ and its reparametrised counterpart in the Einstein frame $\tilde{A}$ are equivalent. This behaviour is represented by the dashed red line in Figure~\ref{Fig:ALegendLogPlot}.
During inflation (at high field values), the $\tilde{A}$ field is defined as $\log(\Omega^2)$ as discussed in Eq.~(\ref{eq:def-A}), which is shown by the dashed green curve in Figure~\ref{Fig:ALegendLogPlot}. 
The low and high field regions are separated by $\tilde{A}_{cr}$, at the intersection of the dashed red and green curves where $\tilde{h}_1 = \log(\Omega^2)$ and is calculated to be
\be  
\label{eq:reheating-field-approx}
\tilde{h}_1 = \log \left(1+{\hat{\xi}}_4 \, {\tilde{h}_1}^2 \right) \,
\approx \, {\hat{\xi}}_4 \, {\tilde{h}_1}^2
\quad \Rightarrow \quad \tilde{h}_1 = \frac{1}{\hat{\xi}_4 } \, \equiv \, \tilde{A}_{cr}\,,
\ee
using the assumption of $\hat{\xi}_4 \gg 1$. 

Another important intersection is the $\tilde{A} \sim 1$ point, above which inflation occurs.
In the intermediate region where 
$1> \tilde{A}>\tilde{A}_{cr} = 1/\hat{\xi}_4 $, the field $\tilde{A}$ can be approximated as ${\hat{\xi}}_4 {\tilde{h}_1}^2$ which is represented by the dashed magenta curve in Figure~\ref{Fig:ALegendLogPlot}. 
In this region, 
which is relevant for reheating,
the inflationary potential in Eq.~(\ref{eq:full-pot-T1T4}) can be approximated by a quadratic potential
\be 
\label{eq:pot-oscillation}
\tilde{V}
\, \approx \,
 \biggl(\frac{M_{pl}^2}{2 \,|\xi_4|} \biggr)^2
X(\theta_1,\theta_4)  \,
\tilde{A}^2 
~ \equiv ~
\frac{1}{2} \,\omega^2\,\tilde{A}^2
\qquad 
\mbox{with}
\quad  
\omega = \frac{M_{pl}^2\,  \sqrt{X(\theta_1,\theta_4)}}{\sqrt{2} \,|\xi_4|}\,,
\ee 
which is a simple harmonic oscillator potential in which the inflaton oscillates rapidly with frequency $\omega$ which could be thought of as the ``inflaton mass''.
Since the effective inflaton mass is non-zero in this region, the exponential expansion of the universe proceeds as in the matter domination era where the Friedman
equation can be written as
\be 
\label{eq:Friedman}
3\, H^2(t) =  
\frac{1}{2}\dot{\tilde{A}}^2(t) + \frac{1}{2}\frac{\omega^2}{{M_{pl}}^2}\tilde{A}^2(t),
\qquad
\mbox{where}
\quad 
H(t)=\frac{\dot a(t)}{a(t)}=\frac{2}{3t}
\quad
\mbox{and}
\quad
a\propto t^{2/3}\,,
\ee
where $t$ is the physical time, $H$ is the Hubble parameter, $a$ is the scale factor and $\omega$ acting as a mass parameter for this oscillatory phase.
This equation can be solved for $\omega\gg H$ (when
the change of the scale factor is small during one oscillation) as
\be
\label{eq:background-approx} 
\tilde{A}=\tilde{A}_0(t)\cos (\omega t)\,,
\ee
where $\tilde{A}_0(t)$ is the amplitude of the background inflaton
field oscillations, which decreases with time due to particle creation and the expansion of the universe where
\be  
\tilde{A}_0 = \sqrt{\frac{2}{3}}\frac{A_0}{M_{pl}} =
\frac{8\, |\xi_4|}{\sqrt{3} \,M_{pl}\, \sqrt{X(\theta_1,\theta_4)}}\, \frac{1}{t}  \,.
\ee 
The reheating phase ends at time $t_{cr}$ when the amplitude of the oscillations $\tilde{A}_0$ crosses $\tilde{A}_{cr}$ which gives us the crossing time as
\be  
\label{eq:t-critical}
t_{cr}= \frac{8\sqrt{2}\,M_{pl} \, |\xi_4| \beta_2(c_{\theta_4} +\beta_1 s_{\theta_4})}{\sqrt{3}\, \omega} =
\frac{4\sqrt{2}\,M_{pl} \, \hat{\xi}_4}{\sqrt{3}\, \omega}\,.
\ee 
At later times, when $\tilde{A} <\tilde{A} _{cr}$, the universe enters the radiation-dominated era. The potential for the inflaton field no longer contains an essential mass parameter. 
The energy of the inflaton zero mode is drained by the
creation of SM-Higgs and gauge bosons through their direct coupling to the inflaton filed.
These couplings are relatively large and lead to a rapid energy transfer from the coherent
oscillations of the inflaton to relativistic SM particles.  If this energy conversion is
instantaneous, a lower bound on the reheating temperature is estimated to be 
$T_{reh} >1.5 \times 10^{13}$ GeV \cite{Bezrukov:2007ep,Bezrukov:2008cq}.

\subsection{Inflaton decay}

As discussed before, the inflaton decays into the $W$, $Z$ and the SM-Higgs bosons through its direct couplings. 
In the background approximation in Eq.~(\ref{eq:background-approx}) the inflaton field acts as an external source of these SM fields with ``varying-with-time masses''. Therefore, although 
these particles do not have a physical mass at this time, one can define an effective mass arising from inflaton oscillations for them. 
Using the same assumption as before, 
where $\omega \gg H$ and the amplitude is constant over one oscillation period, we define effective mass terms for $W$, $Z$ and the Higgs bosons. 
Recall that during the reheating phase, $h_1^2$ can be approximated as $M^2_{pl} \, \tilde{A}/ \hat{\xi}_4$ as shown in Eq.~(\ref{eq:reheating-field-approx}).

The coupling of the inflaton to $W$ and $Z$ bosons is $\frac{1}{4}g_2^2 \,h_1^2W^2$ and $\frac{1}{8}(g_1^2+g_2^2) \,h_1^2Z^2$, respectively. 
Written in terms of $\tilde{A}$, the effective masses for the $W$ and $Z$ bosons are calculated to be:
\be
\label{eq:mW-mZ}
m_W^2 = \frac{g_2^2\, {M^2_{pl}}}{4\,\hat{\xi}_4} \, \tilde{A},
\qquad 
m_Z^2  =
\frac{(g_1^2+g_2^2)\, M^2_{pl}}{4\, \hat{\xi}_4} \, \tilde{A}\,,
\ee
where $g_1$ and  $g_2$ are the $U(1)$ and $SU(2)$ gauge couplings, respectively, and $\hat{\xi}_4$ is defined in Eq.~(\ref{eq:shorthand}). 
The inflaton coupling to Higgs is through the potential in Eq.~(\ref{eq:V0-3HDM}) expanded in terms of all fields in Eq.~(\ref{eq:explicit-fields}), which allows us to define an effective mass term for the SM-like Higgs boson, $h_3$, as
\be 
\label{eq:mhSq}
m_{h_3}^2 = \frac{\lambda_{123}}{2}  h_1^2  = 
\frac{\lambda_{123}}{2} \frac{M^2_{pl}}{\hat{\xi}_4} \, \tilde{A}\,,
\ee
where
\be
\label{eq:lambda123}
\lambda_{123} = \beta_2^2(\lambda_{23}+\lambda'_{23})+
(1+\beta_1^2)(\lambda_{31}+\lambda'_{31})+
2 \beta_2^2 \lambda_2 c_{\theta_2} -
2(\beta_1^2-1)\lambda_3 c_{\theta_3} -
4\beta_1 \lambda_3 s_{\theta_3}\,,
\ee
and $\theta_i$ being the phase of the parameter $\lambda_i$ as described in Eq.~(\ref{eq:complex-params}).
Note that the masses in Eqs.(\ref{eq:mW-mZ}) and (\ref{eq:mhSq}) are not the conventional masses obtained from spontaneous symmetry breaking. Instead they are effective masses defined based on the interactions of these fields with the inflaton written in terms of the reparametrised field $\tilde{A}$.

Due to the large value of the weak couplings, $g_1$ and $g_2$ of the $W,\,Z$ bosons to the inflaton, they are heavy and non-relativistic. If $\lambda_{123}$ is large, the Higgs boson will also be produced non-relativistically. 
Therefore, their production does not change the equation of state from the non-relativistic matter domination to radiation domination. 
This transition occurs later on with the creation of the relativistic secondary particles, i.e. light fermions,  as a result of the decay or scattering of the heavy particles, the Higgs and $W,\,Z$ bosons. 

As long as the inflaton mass, $\omega$, is smaller than
the gauge/scalar bosons in Eqs.~(\ref{eq:mW-mZ}, \ref{eq:mhSq})  for $\tilde{A} > \tilde{A}_{cr}$, creation of the bosons is
possible only when the inflaton field crosses zero
(when $\tilde{A}(t) < \tilde{A}_{cr}$).  During each zero crossing some gauge/scalar bosons are created.  
In the beginning, when the number densities of the produced $W,Z$ and Higgs bosons, $n_{W}, n_{Z}$ and $n_h$ respectively, is low the creation rate is constant. As a result, the created bosons are non-relativistic and decay into light SM fermions (which are relativistic).  
However, the decay rate decreases with time as the amplitude of the inflaton oscillations decreases. No significant energy transfer from the inflaton to radiation has taken place so far.
As the decay rate becomes smaller than the production rate, generation of the particles, which is enhanced by the stochastic parametric resonance, accelerates and raises the concentration of the gauge/scalar bosons. The energy transfer into the light SM fermions proceeds now mainly via $WW/ZZ\to f\bar{f}$ annihilation (their decays become sub-dominant channels of fermion production) while Higgs can only produce fermions through decays.  
The annihilation process rapidly transfers all the energy into radiation, resulting in the transition from
the matter domination expansion with $a\propto t^{2/3}$ to the radiation domination expansion with $a\propto t^{1/2}$.

The production of $W$ and Higgs bosons in the linear and resonance regions are \cite{Bezrukov:2008ut,GarciaBellido:2008ab,Repond:2016sol}:
\be 
\label{eq:W-production}
\frac{d(n_W a^3)}{dt}=
\left\{\begin{array}{c} 
\frac{P}{2\pi^3}\omega K_1^3 a^3, 
~~~~~~~~~ \mbox{(linear)}, 
\\ \\ 
2\,a^3\,\omega Q\,n_W, 
~~~~~~~\mbox{(resonance)},
\end{array}
\right.
\qquad \qquad
\frac{d(n_ha^3)}{dt}=\left\{
\begin{array}{c} 
\frac{P}{2\pi^3}\omega K_2^3 a^3,
~~~~~~~~~ \mbox{(linear)}, 
\\ \\ 
2 a^3 \omega Q n_h,
~~~~~~~\mbox{(resonance)},
\end{array}
\right.
\ee
where $P$ and $Q$ are numerical factors with $P\approx 0.0455$ and $Q\approx 0.045$. The created particles are
essentially non-relativistic.  For concentrations of other gauge
bosons we have the obvious relations $n_{W^+}=n_{W^-}$,
$n_Z=n_{W^+}/\cos^2\theta_W$, where $\theta_W$ is the weak mixing angle.
$K_1$ and $K_2$ have dimensions of energy and are dependent on the respective mass terms with:
\bea 
K_1^3&=& \omega \, m^2_{W}(t_i) = \frac{g_2^2 \, M_{pl}^4}{2\sqrt{2}\, {\hat{\xi}_4}^2}\, \beta_2(c_{\theta_4} +\beta_1 s_{\theta_4}) \sqrt{X(\theta_1,\theta_4)}
\, \tilde{A}_0(t_i) , \\
K_2^3 &=&  \omega \, m^2_{h_3}(t_i) =\frac{\lambda_{123} \,M_{pl}^4}{\sqrt{2} \,{\hat{\xi}_4}^2}\, \beta_2(c_{\theta_4} +\beta_1 s_{\theta_4}) \sqrt{X(\theta_1,\theta_4)}
\, \tilde{A}_0(t_i) ,
\eea
where $t_i$ is the instant when the inflaton field crosses zero, i.e. $ \tilde{A}(t_i)=0$. 
Note that the inflaton can decay into $W,Z$ and Higgs bosons only in the vicinity of this point, when its effective mass $\omega$, is much larger than those of the $W,Z$ and Higgs bosons.

\section{Scalar asymmetries}
\label{sec:reheating}
Here we briefly discuss how the CP asymmetry
originating from the non-minimal coupling, is transferred to the SM
degrees of freedom.
For this discussion, let's focus on the neutral components of the $\phi_1$ doublets acquiring an initial non-vanishing expectation value at the exit from
inflaton. We write the field fluctuations around the initial conditions as
\be
\label{eq:fluctuations}
\left\{\begin{array}{c}
\phi_1 \to \phi_1 - a_1 e^{i \,\alpha} \\
\phi_1^\dagger \to \phi_1^* - a_1 e^{-i \,\alpha}
\end{array}
\right. 
,
\qquad
\left\{\begin{array}{c}
\phi_2 \to \phi_2 - a_2\\
\phi_2^\dagger \to \phi_2^* - a_2
\end{array}
\right. 
,
\qquad
\left\{\begin{array}{c}
\phi_3 \to \phi_3 - a_3\\
\phi_3^\dagger \to \phi_3^* - a_3
\end{array}
\right.
.
\ee
The phase $\alpha$ here is related to the CP-violating phases of inflation.
Note that at the end of inflation the $h_1$ field has taken a value according to Eq.~(\ref{eq:hi&hf}) which is dependant on the inflationary dynamics, namely $\theta_4$, $\beta_1$ and $\beta_2$ which are dependant on $\theta_1$.
Since $h_1$ is the real part of the complex field $\phi_1$, its value is what feeds the
$a_1 \cos\alpha$ component of fluctuations in
Eq.~(\ref{eq:fluctuations}). The imaginary part of $\phi_1$, represented by
$\eta_1$, takes a value proportional to $h_1$ and feeds the $a_1 \sin\alpha$ component of the field fluctuations.
Recall that one can obtain the values of $\beta_1$ and $\beta_2$ for any given value of $\theta_1$ and $\theta_4$.
However, to keep the present discussion more transparent, we retain
a generic phase $\alpha$ here.

To discuss the consequences of this complex phase, we now
assume instant reheating. Since the field $\phi_3$ is light with respect to the inflaton
degrees of freedom, we expect the latter to quickly decay to
$\phi_3$. The asymmetry arising from the values of the fields
in Eq.~(\ref{eq:fluctuations}) will manifest in creation of unequal number
of $\phi_3$ and $\phi_3^\ast$ quanta as follows.

From the potential in Eq.~(\ref{eq:V0-3HDM}), the couplings
contributing to the decays of
\be 
\left\{\begin{array}{c}
\phi_1 \to \phi_3 \phi_3 \; \propto \;2 a_1 \lambda_3 \, e^{i (\alpha +\theta_3)} \\
\phi_1^* \to \phi_3^* \phi_3^* \; \propto \; 2 a_1 \lambda_3 \, e^{-i (\alpha +\theta_3)}
\end{array} \right.
,
 \qquad
\left\{\begin{array}{c}
\phi_2 \to \phi_3 \phi_3 \; \propto \;2 a_2 \lambda_2 \, e^{i \theta_2} \\
\phi_2^* \to \phi_3^* \phi_3^* \; \propto \; 2 a_2 \lambda_2 \, e^{-i \theta_2}
\end{array} \right.
.
\ee 
Such decay processes are CP-violating and result in unequal number of $\phi_3$ and $\phi_3^*$ states. Consequently, the relative asymmetries $A^1_{CP}$ and $A^2_{CP}$ in the decay rates are
\be
A^1_{CP} \; \sim \; 8 \,a_1^2\, \lambda_3^2 \,\sin2(\alpha+\theta_3)\,,
\qquad
A^2_{CP} \; \sim \; 8\, a_2^2\, \lambda_2^2\, \sin2\theta_2\,.
\ee
This asymmetry in the scalar sector is then transferred to the fermion sector through the couplings of the Higgs field ($h_3$ contained in the $\phi_3$ doublet) with the fermions, as discussed in the previous section.
For example, assuming the existence of right-handed neutrinos, the Yukawa interactions
between neutrinos and $\phi_3$ will generate an asymmetry between $\nu_L$ and $\bar{\nu}_R$,
which would be further translated into baryon asymmetry by the electroweak sphalerons.

\section{Conclusion and outlook}
\label{sec:conclusions}
Scalar fields which have non-minimal couplings to gravity are well-motivated inflaton candidates. Paradigmatic examples are the Higgs-inflation \cite{Bezrukov:2007ep}
and $s$-inflation models \cite{Enqvist:2014zqa}.
In this paper we have considered a scenario where several non-minimally coupled scalars
contribute to the inflationary dynamics. In particular we investigated a model where
these scalars are electroweak doublets and therefore generalize the Higgs inflation.
We focused on a setting where the dominant non-minimal coupling is allowed to be complex
and investigated the effect that this would have on CP-violation in our universe.
We determined the inflationary dynamics in the regime where the model essentially
conforms to the predictions of single field inflation. The essential difference is
that the inflaton obtains a non-zero phase representing possible source of CP-violation
for subsequent post-inflationary evolution.
At the end of inflation, the inflaton particle which is naturally assumed to have couplings with the SM Higgs, dumps its energy into the SM particle bath through the process of
reheating, which populates the universe with the SM particles. We sketched how
the complex value of the inflaton field leads to an asymmetry in the scalar sector decays,
and how this asymmetry will further be transmitted to the fermion sector.
A more detailed analysis of our framework, including multi-field dynamics during inflation,
further details of reheating and subsequent particle decays and their
effects on the generation of baryon asymmetry are covered in our upcoming publication.

\section*{Acknowledgements}
The author acknowledges financial support from Academy of Finland projects ``Particle cosmology and gravitational waves'' No. 320123
and ``Particle cosmology beyond the Standard Model'' No. 310130, and 
would like to thank the organisers of the BSM-2021 conference for creating the opportunity for lively scientific discussions.

\bibliographystyle{unsrt}

\begin{thebibliography}{99}

\bibitem{Aad:2012tfa}
G.~Aad \textit{et al.} [ATLAS],
Phys. Lett. B \textbf{716}, 1-29 (2012)
doi:10.1016/j.physletb.2012.08.020
[arXiv:1207.7214 [hep-ex]].

\bibitem{Chatrchyan:2012ufa}
S.~Chatrchyan \textit{et al.} [CMS],
Phys. Lett. B \textbf{716}, 30-61 (2012)
doi:10.1016/j.physletb.2012.08.021
[arXiv:1207.7235 [hep-ex]].

\bibitem{Ade:2015xua}
P.~A.~R.~Ade \textit{et al.} [Planck],
Astron. Astrophys. \textbf{594}, A13 (2016)
doi:10.1051/0004-6361/201525830
[arXiv:1502.01589 [astro-ph.CO]].

\bibitem{Morrissey:2012db}
D.~E.~Morrissey and M.~J.~Ramsey-Musolf,
New J. Phys. \textbf{14}, 125003 (2012)
doi:10.1088/1367-2630/14/12/125003
[arXiv:1206.2942 [hep-ph]].

\bibitem{Gavela:1993ts}
M.~B.~Gavela, P.~Hernandez, J.~Orloff and O.~Pene,
Mod. Phys. Lett. A \textbf{9}, 795-810 (1994)
doi:10.1142/S0217732394000629
[arXiv:hep-ph/9312215 [hep-ph]].

\bibitem{Huet:1994jb}
P.~Huet and E.~Sather,
Phys. Rev. D \textbf{51}, 379-394 (1995)
doi:10.1103/PhysRevD.51.379
[arXiv:hep-ph/9404302 [hep-ph]].

\bibitem{Gavela:1994dt}
M.~B.~Gavela, P.~Hernandez, J.~Orloff, O.~Pene and C.~Quimbay,
Nucl. Phys. B \textbf{430}, 382-426 (1994)
doi:10.1016/0550-3213(94)00410-2
[arXiv:hep-ph/9406289 [hep-ph]].

\bibitem{Kajantie:1995kf}
K.~Kajantie, M.~Laine, K.~Rummukainen and M.~E.~Shaposhnikov,
Nucl. Phys. B \textbf{466}, 189-258 (1996)
doi:10.1016/0550-3213(96)00052-1
[arXiv:hep-lat/9510020 [hep-lat]].

\bibitem{Hawking:1982cz}
S.~W.~Hawking,
Phys. Lett. B \textbf{115}, 295 (1982)
doi:10.1016/0370-2693(82)90373-2

\bibitem{Starobinsky:1982ee}
A.~A.~Starobinsky,
Phys. Lett. B \textbf{117}, 175-178 (1982)
doi:10.1016/0370-2693(82)90541-X

\bibitem{Sasaki:1986hm}
M.~Sasaki,
Prog. Theor. Phys. \textbf{76}, 1036 (1986)
doi:10.1143/PTP.76.1036

\bibitem{Mukhanov:1988jd}
V.~F.~Mukhanov,
Sov. Phys. JETP \textbf{67}, 1297-1302 (1988)

\bibitem{Bezrukov:2007ep}
F.~L.~Bezrukov and M.~Shaposhnikov,
Phys. Lett. B \textbf{659}, 703-706 (2008)
doi:10.1016/j.physletb.2007.11.072
[arXiv:0710.3755 [hep-th]].

\bibitem{Lerner:2009xg}
R.~N.~Lerner and J.~McDonald,
Phys. Rev. D \textbf{80}, 123507 (2009)
doi:10.1103/PhysRevD.80.123507
[arXiv:0909.0520 [hep-ph]].

\bibitem{Enqvist:2014zqa}
K.~Enqvist, S.~Nurmi, T.~Tenkanen and K.~Tuominen,
JCAP \textbf{08}, 035 (2014)
doi:10.1088/1475-7516/2014/08/035
[arXiv:1407.0659 [astro-ph.CO]].

\bibitem{Gong:2012ri}
J.~O.~Gong, H.~M.~Lee and S.~K.~Kang,
JHEP \textbf{04}, 128 (2012)
doi:10.1007/JHEP04(2012)128
[arXiv:1202.0288 [hep-ph]].

\bibitem{Choubey:2017hsq}
S.~Choubey and A.~Kumar,
JHEP \textbf{11}, 080 (2017)
doi:10.1007/JHEP11(2017)080
[arXiv:1707.06587 [hep-ph]].

\bibitem{Englert:2011yb}
C.~Englert, T.~Plehn, D.~Zerwas and P.~M.~Zerwas,
Phys. Lett. B \textbf{703}, 298-305 (2011)
doi:10.1016/j.physletb.2011.08.002
[arXiv:1106.3097 [hep-ph]].

\bibitem{Branco:2011iw}
G.~C.~Branco, P.~M.~Ferreira, L.~Lavoura, M.~N.~Rebelo, M.~Sher and J.~P.~Silva,
Phys. Rept. \textbf{516}, 1-102 (2012)
doi:10.1016/j.physrep.2012.02.002
[arXiv:1106.0034 [hep-ph]].

\bibitem{Keus:2016orl}
V.~Keus,
PoS \textbf{CHARGED2016}, 017 (2016)
doi:10.22323/1.286.0017
[arXiv:1612.03629 [hep-ph]].

\bibitem{Cordero-Cid:2016krd}
A.~Cordero-Cid, J.~Hern\'andez-S\'anchez, V.~Keus, S.~F.~King, S.~Moretti, D.~Rojas and D.~Soko\l{}owska,
JHEP \textbf{12}, 014 (2016)
doi:10.1007/JHEP12(2016)014
[arXiv:1608.01673 [hep-ph]].

\bibitem{Cordero:2017owj}
A.~Cordero, J.~Hernandez-Sanchez, V.~Keus, S.~F.~King, S.~Moretti, D.~Rojas and D.~Sokolowska,
JHEP \textbf{05}, 030 (2018)
doi:10.1007/JHEP05(2018)030
[arXiv:1712.09598 [hep-ph]].

\bibitem{Cordero-Cid:2018man}
A.~Cordero-Cid, J.~Hern\'andez-S\'anchez, V.~Keus, S.~Moretti, D.~Rojas and D.~Soko\l{}owska,
Eur. Phys. J. C \textbf{80}, no.2, 135 (2020)
doi:10.1140/epjc/s10052-020-7689-0
[arXiv:1812.00820 [hep-ph]].

\bibitem{Cordero-Cid:2020yba}
A.~Cordero-Cid, J.~Hern\'andez-S\'anchez, V.~Keus, S.~Moretti, D.~Rojas-Ciofalo and D.~Soko\l{}owska,
Phys. Rev. D \textbf{101}, no.9, 095023 (2020)
doi:10.1103/PhysRevD.101.095023
[arXiv:2002.04616 [hep-ph]].

\bibitem{Keus:2020ooy}
V.~Keus,
PoS \textbf{CORFU2019}, 059 (2020)
doi:10.22323/1.376.0059
[arXiv:2003.02141 [hep-ph]].

\bibitem{Keus:2019szx}
V.~Keus,
Phys. Rev. D \textbf{101}, no.7, 073007 (2020)
doi:10.1103/PhysRevD.101.073007
[arXiv:1909.09234 [hep-ph]].

\bibitem{Hernandez-Sanchez:2020aop}
J.~Hernandez-Sanchez, V.~Keus, S.~Moretti, D.~Rojas-Ciofalo and D.~Sokolowska,
[arXiv:2012.11621 [hep-ph]].

\bibitem{Ahriche:2015mea}
A.~Ahriche, G.~Faisel, S.~Y.~Ho, S.~Nasri and J.~Tandean,
Phys. Rev. D \textbf{92}, no.3, 035020 (2015)
doi:10.1103/PhysRevD.92.035020
[arXiv:1501.06605 [hep-ph]].

\bibitem{Ivanov:2011ae}
I.~P.~Ivanov, V.~Keus and E.~Vdovin,
J. Phys. A \textbf{45}, 215201 (2012)
doi:10.1088/1751-8113/45/21/215201
[arXiv:1112.1660 [math-ph]].

\bibitem{Keus:2013hya}
V.~Keus, S.~F.~King and S.~Moretti,
JHEP \textbf{01}, 052 (2014)
doi:10.1007/JHEP01(2014)052
[arXiv:1310.8253 [hep-ph]].

\bibitem{Kaiser:2010ps}
D.~I.~Kaiser,
Phys. Rev. D \textbf{81}, 084044 (2010)
doi:10.1103/PhysRevD.81.084044
[arXiv:1003.1159 [gr-qc]].

\bibitem{Keus:2021dti}
V.~Keus and K.~Tuominen,
[arXiv:2102.07777 [hep-ph]].

\bibitem{Ade:2015lrj}
P.~A.~R.~Ade \textit{et al.} [Planck],
Astron. Astrophys. \textbf{594}, A20 (2016)
doi:10.1051/0004-6361/201525898
[arXiv:1502.02114 [astro-ph.CO]].

\bibitem{Komatsu:2010fb}
E.~Komatsu \textit{et al.} [WMAP],
Astrophys. J. Suppl. \textbf{192}, 18 (2011)
doi:10.1088/0067-0049/192/2/18
[arXiv:1001.4538 [astro-ph.CO]].

\bibitem{Linde:1981mu}
A.~D.~Linde,
Phys. Lett. B \textbf{108}, 389-393 (1982)
doi:10.1016/0370-2693(82)91219-9

\bibitem{Bezrukov:2008cq}
F.~L.~Bezrukov,
[arXiv:0805.2236 [hep-ph]].

\bibitem{Bezrukov:2008ut}
F.~Bezrukov, D.~Gorbunov and M.~Shaposhnikov,
JCAP \textbf{06}, 029 (2009)
doi:10.1088/1475-7516/2009/06/029
[arXiv:0812.3622 [hep-ph]].

\bibitem{GarciaBellido:2008ab}
J.~Garcia-Bellido, D.~G.~Figueroa and J.~Rubio,
Phys. Rev. D \textbf{79}, 063531 (2009)
doi:10.1103/PhysRevD.79.063531
[arXiv:0812.4624 [hep-ph]].

\bibitem{Repond:2016sol}
J.~Repond and J.~Rubio,
JCAP \textbf{07}, 043 (2016)
doi:10.1088/1475-7516/2016/07/043
[arXiv:1604.08238 [astro-ph.CO]].


\end{thebibliography}

\end{document}